\documentclass[lettersize,journal]{IEEEtran}
\usepackage{amsmath,amsfonts}
\usepackage{algorithmic}
\usepackage{amsmath}
\usepackage{algorithm}
\usepackage{array}
\usepackage[caption=false,font=footnotesize,labelfont=rm,textfont=rm]{subfig}
\usepackage{textcomp}
\usepackage{stfloats}
\usepackage{url}
\usepackage{verbatim}
\usepackage{graphicx} 
\usepackage{cite}
\usepackage{mathtools}
\usepackage{amssymb}
\usepackage{upgreek}
\usepackage{subfig}
\usepackage{hyperref}
\usepackage{graphicx}
\usepackage{epstopdf}
\usepackage[dvipsnames]{xcolor}
\begin{document}
\title{Near-Field Codebook-Based 3D Spherical Channel Estimation\\ for UCA XL-MIMO Systems}

\author{Chenliang Yang, Guangchi Zhang, Miao Cui, Qingqing Wu,~\IEEEmembership{Senior Member,~IEEE}, and Yong Zeng,~\IEEEmembership{Fellow,~IEEE}
        %% <-this % stops a space
\thanks{The work was supported in part by the Guangdong Basic and Applied Basic Research Foundation under Grants 2023A1515011980 and 2023A1515140003, in part by the BD+ Project of Jiangxi Institute of CMI under Grant 2024JXRH0Y02, and in part by the Natural Science Foundation for Distinguished Young Scholars of Jiangsu under Grant BK20240070.

C. Yang, G. Zhang, and M. Cui are with the School of Information Engineering, Guangdong University of Technology, Guangzhou 510006, China (e-mail: clyang2233@163.com; gczhang@gdut.edu.cn; cuimiao@gdut.edu.cn).
Q. Wu is with the Department of Electronic Engineering, Shanghai Jiao Tong University, Shanghai 200240, China (e-mail: qingqingwu@sjtu.edu.cn). % <-this % stops a space
Y. Zeng is with the National Mobile Communications Research Laboratory, Southeast University, Nanjing 210096, China, also with the Purple Mountain Laboratories, Nanjing 211111, China (e-mail: yong\_zeng@seu.edu.cn).
(Corresponding authors: G. Zhang; M. Cui.) }}

%% The paper headers
\markboth{}%
{}

\IEEEpubid{}
%% Remember, if you use this you must call \IEEEpubidadjcol in the second
%% column for its text to clear the IEEEpubid mark.

\maketitle
\begin{abstract}
Extremely large-scale multiple input multiple output (XL-MIMO), a key technology for 6G communications, faces challenges in near-field channel estimation due to spherical wavefronts and the need for three-dimensional (3D) spatial characterization, particularly with uniform circular arrays (UCAs). This letter proposes a spherical-domain simultaneous orthogonal matching pursuit (S-SOMP) based scheme tailored for near-field 3D channel estimation in UCA-equipped XL-MIMO systems. We establish a sparse channel representation based on the near-field spherical wave model. Then, a novel spherical-domain transform matrix codebook is designed via joint discrete sampling of distance, azimuth, and elevation parameters, leveraging analytical approximations to ensure low correlation between steering vectors. This structured codebook enables accurate sparse signal recovery using the S-SOMP algorithm for efficient joint estimation of channel path gains, spatial angles, and distances. Simulation results demonstrate significant channel estimation accuracy improvements compared to existing benchmarks.
\end{abstract}


\section{Introduction}
\IEEEPARstart{E}{xtremely} large-scale multiple input multiple output (XL-MIMO) is expected to be a key enabling technology for 6G wireless communications, offering substantial gains in spatial resolution and spectral efficiency through the deployment of massive antenna arrays \cite{re1,re1_1}. However, the increased aperture size makes near-field propagation effects non-negligible, invalidating the traditional far-field assumption with uniform plane wave (UPW) models. While much XL-MIMO research focuses on uniform linear arrays (ULAs), their near-field coverage can be limited, particularly for users at large angles relative to the array broadside \cite{re2}. Furthermore, the ULA geometry struggles to capture three-dimensional (3D) near-field channel characteristics, which depend on distance, azimuth, and elevation, as it primarily resolves two-dimensional (2D) parameters. In contrast, uniform circular arrays (UCAs), due to their rotational symmetry, provide uniform near-field coverage across all angles \cite{re3}. UCAs are well-suited for characterizing 3D near-field spatial propagation, enabling joint modeling of distance, azimuth, and elevation angles \cite{re4}. Therefore, investigating near-field XL-MIMO systems with UCAs is an important research direction.

Realizing the benefits of XL-MIMO systems hinges on acquiring accurate channel state information (CSI)\cite{re5}. However, the large number of antennas drastically increases pilot overhead and computational complexity for channel estimation. compressed sensing (CS)-based algorithms offer promising solutions \cite{re6,re7}. For instance, \cite{re6} addresses far-field scenarios by exploiting angular-domain sparsity through the spatial Fourier transform, and employs the simultaneous orthogonal matching pursuit (SOMP) algorithm to achieve sparse recovery of multi-subcarrier channels.. Specifically for near-field scenarios, \cite{re7} developed a polar-domain sparse channel model combined with SOMP to improve estimation performance. Nevertheless, these methods predominantly use ULA models and often assume coplanar users, neglecting the elevation dimension crucial for 3D spatial channels. This limits their applicability in realistic environments. While UCAs can capture 3D near-field information, jointly estimating coupled distance, azimuth, and elevation parameters poses significant challenges, particularly for achieving effective sparse representation. Systematic solutions are still missing.

Motivated by these challenges, this letter proposes a spherical-domain SOMP (S-SOMP) method for near-field channel estimation in UCA-based XL-MIMO systems. Our main contributions are:
\begin{itemize}
  \vspace{-0.1cm}
    \item We establish a spherical-domain sparse channel model specifically tailored for 3D near-field propagation with UCAs. This model accurately captures the inherent coupling between distance, azimuth, and elevation parameters, overcoming the limitations of conventional 2D representations in such scenarios.
    \item We propose a novel method for designing the spherical-domain transform codebook, which is critical for effective sparse representation. Unlike heuristic grid sampling, our principled approach determines sampling intervals for $r, \theta, \phi$ using analytical approximations derived by controlling steering vector mutual coherence (targeting Bessel function zeros), significantly enhancing subsequent CS recovery performance.
    \item We develop a complete and efficient S-SOMP channel estimation scheme by integrating the tailored, low-coherence spherical-domain codebook with the SOMP algorithm. This integration effectively leverages the joint sparsity structure across subcarriers, providing a robust solution for accurate 3D near-field channel estimation in UCA XL-MIMO systems.
    \vspace{-0.1cm}
\end{itemize}
Simulation results validate that the proposed S-SOMP algorithm, using the novel codebook design, achieves significantly improved estimation accuracy compared to conventional angular-domain and polar-domain methods, especially for joint estimation of multi-dimensional parameters in 3D UCA near-field environments.

\section{System Model}
\begin{figure}[!t]
  \centering
  \includegraphics[width=2in]{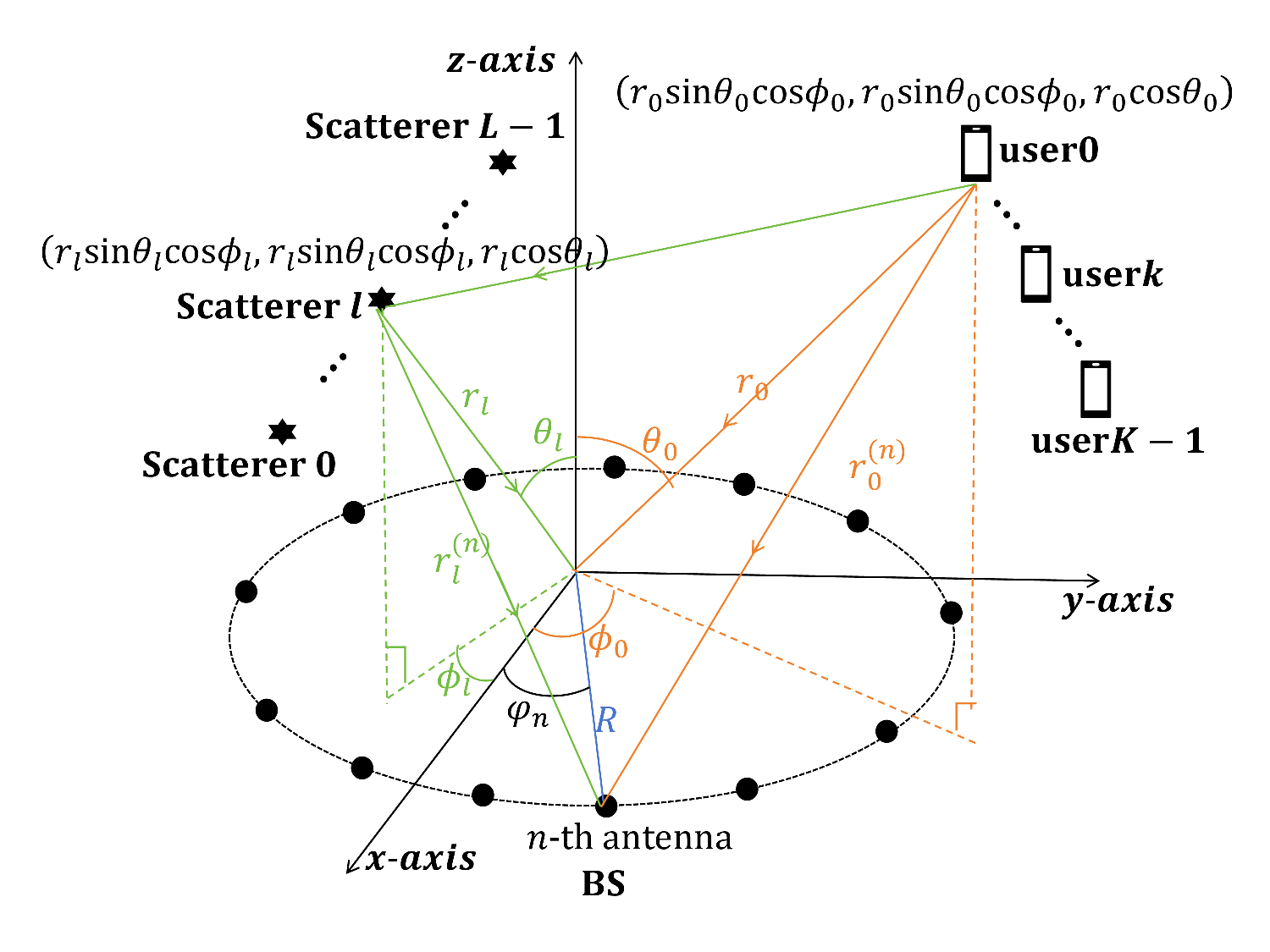}
  \caption{An XL-MIMO OFDM system with UCA.}
  \vspace{-0.25cm}
  \label{figure1}
\end{figure}

We consider an XL-MIMO orthogonal frequency division multiplexing (OFDM) system, as shown in Fig. \ref{figure1}. The base station (BS) has a UCA with $N$ antennas serving $K$ single-antenna users.
A hybrid precoding architecture is employed at the BS with $N_{\mathrm{RF}}$ radio frequency (RF) chains ($N_\mathrm{RF} \ll N$). We use a 3D cartesian coordinate system centered at the UCA, with the UCA in the $x$-$y$ plane. The coordinate of the $n$-th antenna is $(R \cos \psi_n, R \sin \psi_n, 0)$, $n \in \{0, \ldots, N-1\}$, where $R$ is the UCA radius and $\psi_n = \frac{2\pi n}{N}$ is the angular position (azimuth) of the $n$-th antenna. The system uses $M$ sub-carriers, and the frequency of the $m$-th sub-carrier is $f_m = f_{\rm c} + \frac{(2m - M)B}{2M}$, for $m \in \{ 1, \ldots, M \}$. Here, $B$ is the total bandwidth and $f_{\rm c}$ is the center carrier frequency. The nominal adjacent antenna spacing is $d = \lambda/2$, where $\lambda = \mathrm{c}/f_\mathrm{c}$ is the center carrier wavelength, and thus $R = \frac{d}{2} \mathrm{sin}(\frac{\pi}{N})$.

We assume the $K$ users transmit mutually orthogonal pilot sequences to the BS. which allows the channel for each user to be estimated independently. Without loss of generality, we therefore focus on an arbitrary user.
Specifically, the user transmits $P$ OFDM pilot signals over $P$ time slots, with $x_{m,p}$ denoting the signal over sub-carrier $m$ during time slot $p$, $p=1, \ldots, P$. We assume that the channel remains unchanged over $P$ time slots. The received signal at the BS on sub-carrier $m$ in time slot $p$, $\mathbf{y}_{m,p} \in \mathbb{C}^{N_\mathrm{RF} \times 1}$, is
\begin{equation}
  \mathbf{y}_{m,p} = \mathbf{A}_p \mathbf{h}_m x_{m,p} + \mathbf{n}_{m,p},
\end{equation}
where $\mathbf{A}_p \in \mathbb{C}^{N_\mathrm{RF} \times N}$ is the analog combining matrix for time slot $p$, satisfying the constant modulus constraint $|\mathbf{A}_p(i,j)| = 1/\sqrt{N}$. $\mathbf{h}_m \in \mathbb{C}^{N \times 1}$ is the channel vector for sub-carrier $m$. $\mathbf{n}_{m,p} \sim \mathcal{CN}(0, \sigma^2 \mathbf{I}_{N_{\mathrm{RF}}})$ is the additive complex Gaussian noise vector. Without loss of generality, we assume $x_{m,p}=1$ for all $m, p$ \cite{re7}. Aggregating received signals over $P$ time slots for sub-carrier $m$, $\mathbf{y}_m = [\mathbf{y}^\mathrm{T}_{m,1}, \ldots, \mathbf{y}^\mathrm{T}_{m,P}]^{\mathrm{T}}$, gives
\begin{equation}
  \label{eq2}
  \mathbf{y}_m = \mathbf{A} \mathbf{h}_m + \mathbf{n}_m,
\end{equation}
where $\mathbf{A} = [ \mathbf{A}_1^T, \ldots, \mathbf{A}_P^T ]^\mathbf{T} \in \mathbb{C}^{PN_{\mathrm{RF}} \times N}$ is the overall combining matrix and $\mathbf{n}_m = [\mathbf{n}_{m,1}^{\mathrm{T}}, \ldots, \mathbf{n}_{m,P}^{\mathrm{T}}]^{\mathrm{T}} \in \mathbb{C}^{PN_{\mathrm{RF}} \times 1}$ is the aggregated noise.

Due to the large $N$, the Rayleigh distance is extended.
Consequently, the channel requires the near-field spherical wave model, not the far-field plane wave model.
Under the near-field assumption, the channel vector $\mathbf{h}_m$ is a superposition of $L$ paths \cite{re8}:
\begin{equation}
  \label{eq3}
  \mathbf{h}_{m} = \sqrt{\frac{N}{L}} \sum_{l=0}^{L-1} g_{l} e^{-j k_{m} r_{l}} \mathbf{b}(r_l, \theta_{l}, \phi_{l}),
\end{equation}
where $g_l$ is the complex gain of the $l$-th path, $L$ is the path count, and $k_m = \frac{2\pi f_m}{c}$ is the wavenumber for sub-carrier $m$.
$r_l$, $\phi_l$, and $\theta_l$ are the distance, azimuth, and elevation angles for the $l$-th path.
$\mathbf{b}( r_l, \theta_{l}, \phi_{l} ) \in \mathbb{C}^{N \times 1}$ is the UCA near-field steering vector \cite{re4}:
\begin{equation}
  \label{eq4}
  \mathbf{b} ( r_l, \theta_{l}, \phi_{l} ) = \frac { 1 } { \sqrt { N } } \Bigg [ e ^ { - j \frac { 2 \pi } { \lambda } \left( r _ { l } ^ { ( 0 ) } - r _ { l } \right) } , \cdots , e ^ { - j \frac { 2 \pi } { \lambda } \left( r _ { l } ^ { ( N - 1 ) } - r _ { l } \right) } \Bigg ]^\mathrm{T},
\end{equation}
where $r_l^{(n)}$ is the exact distance from the $n$-th antenna to the user/scatterer for path $l$, given by
\begin{equation}
  \label{eq5}
  \begin{split}
  r_l^{(n)} &= \sqrt{(r_l)^2+R^2-2R r_l \sin \theta_l \cos (\phi_l-\psi_n)} \\
  &\approx r_l - R \sin \theta_l \cos (\phi_l-\psi_n) \\
  &\quad + \frac{R^2}{2r_l} \left( 1-\sin^2 \theta_l \cos^2 (\phi_l-\psi_n) \right).
  \end{split}
\end{equation}
The approximation in \eqref{eq5} uses the second-order Taylor expansion $\sqrt{1 + x} \approx 1 + \frac{x}{2} - \frac{x^2}{8}$.

\section{Proposed S-SOMP-Based Channel Estimation Method}
The objective is to estimate the channel matrix $\mathbf{H} \triangleq [\mathbf{h}_1, \ldots, \mathbf{h}_M]$ across all sub-carriers.
Existing near-field schemes often use 2D space (distance, azimuth), inadequately capturing UCA 3D channel characteristics and limiting accuracy.
To address this, we propose an S-SOMP channel estimation method. The method involves three steps:
1) Develop a 3D spherical-domain sparse representation for the UCA near-field channel;
2) Design the corresponding 3D spherical-domain transformation matrix codebook;
3) Use SOMP to estimate the channel via the sparse representation.

\subsection{Sparse Representation of the UCA Near-Field Channels}
Conventional polar-domain representations assume path energy is concentrated only in distance and azimuth.
However, in 3D propagation, elevation $\theta$ significantly influences phase variations (see \eqref{eq4}, \eqref{eq5}).
Thus, single path energy spreads across multiple polar-domain basis vectors, reducing sparsity.
However, the physical channel remains inherently sparse in dominant paths ($L \ll N$, see \eqref{eq3}). This suggests the channel is compressible.

To exploit this in 3D, we propose to represent the near-field channel vector $\mathbf{h}_m$ in a spherical domain:
\begin{equation}
  \label{eq6}
  \mathbf{h}_m = \mathbf{W} \mathbf{h}_m^{\rm S},
\end{equation}
where $\mathbf{h}_m^{\rm S}$ is the spherical-domain sparse vector for sub-carrier $m$. $\mathbf{W} \in \mathbb{C}^{N \times G}$ is the proposed transformation matrix codebook.
It contains $G$ candidate steering vectors $\mathbf{b}_\mathrm{S}(r_g, \theta_g, \phi_g)$ ($g=1, \ldots, G$), each corresponding to a sampled point $(r_g, \theta_g, \phi_g)$ in the 3D space (distance, elevation, azimuth). The expression of $\mathbf{b}_\mathrm{S}(r_g, \theta_g, \phi_g)$ is consistent with \eqref{eq4}. Since this codebook spans the relevant 3D parameter space, we refer to it as the spherical-domain transform matrix codebook. The value of $G$, the total number of columns in the codebook, depends on the sampling strategy detailed next.

\subsection{Design of the Spherical Domain Transformation Matrix Codebook}
The success of CS-based channel estimation relies heavily on the codebook $\mathbf{W}$ properties. For reliable recovery of $\mathbf{h}^S_m$, the mutual coherence between columns of $\mathbf{W}$ must be minimized\cite{re9}.
We design sampling grids for $r, \theta, \phi$ to ensure low correlation between distinct columns $\mathbf{b}_\mathrm{S}(r_{p}, \theta_{p}, \phi_{p})$ and $\mathbf{b}_\mathrm{S}(r_{q}, \theta_{q}, \phi_{q})$ ($p \neq q$). The column correlation is defined as
\begin{equation}
    f(r_{p}, \theta_{p}, \phi_{p}; r_{q}, \theta_{q}, \phi_{q}) \triangleq |\mathbf{b}_\mathrm{S}(r_{p}, \theta_{p}, \phi_{p})^\mathrm{H} \mathbf{b}_\mathrm{S}(r_{q}, \theta_{q}, \phi_{q})|.
\end{equation}
We derive sampling methods for each dimension by controlling this correlation.

\subsubsection{Sampling Method of Elevation Angle}
To determine the sampling interval for elevation angles, we analyze the correlation between two steering vectors corresponding to the same distance and azimuth angle but different elevation angles ($\theta_p \neq \theta_q$). Setting $r_p = r_q = r$ and $\phi_p = \phi_q = \phi$, we can approximate the near-field beamforming gain behavior using far-field approximations for angular resolution \cite{re3}. The column correlation $f(r, \theta_{p}, \phi, r, \theta_{q}, \phi)$ can thus be approximated as
\begin{equation}
  \label{eq9}
  \begin{split}
  &f(r,\theta_p,\phi;r,\theta_q,\phi) = |\mathbf{b}_\mathrm{S}(r,\theta_p,\phi)^\mathrm{H} \mathbf{b}_\mathrm{S}(r,\theta_q,\phi)| \\
  &\approx \left| \mathbf{a}_\mathrm{S}(\theta_p,\phi)^\mathrm{H} \mathbf{a}_\mathrm{S}(\theta_q,\phi) \right|\overset{(a)}= \frac{1}{N} \left| \sum_{n=1}^{N} e^{j\frac{2\pi R}{\lambda}\cos(\phi-\psi_n)\Delta \theta} \right| \\
  &\overset{(b)}{\approx} \frac{1}{N} \left| \sum_{m=-\infty}^{\infty} j^m J_m \left(\frac{2\pi R\Delta\theta}{\lambda}\right)e^{jm\phi} \sum_{n=0}^{N-1} e^{-jm\psi_n} \right| \\
  &\stackrel{(c)}{\approx}\left|J_0\left(\frac{2\pi R\Delta\theta}{\lambda}\right)\right|,
  \end{split}
\end{equation}
where $J_k(\cdot)$ is the $k$-th order Bessel function of the first kind. $\mathbf{a}_\mathrm{S}(\theta, \phi)$ represents the far-field steering vector obtained via first-order Taylor expansion of $\mathbf{b}_\mathrm{S}(r, \theta, \phi)$ with respect to $1/r$. Step $(a)$ uses the far-field approximation with $\Delta \theta = \sin \theta _ { p } - \sin \theta _ { q }$. Step $(b)$ applies the Jacobi-Anger expansion $e^{j \xi \cos \alpha}=\sum_{m=-\infty}^{\infty} j^m J_m(\xi) e^{j m \alpha}$. Step $(c)$ uses the geometric series sum $\sum_{n=0}^{N-1} e^{-jm\psi_n} = N$ if $m$ is a multiple of $N$, and 0 otherwise. For large $N$, the sum is dominated by the $m=0$ term.
To achieve low correlation, we target the zeros of the Bessel function. Setting $\left| J_0 \left( \frac{2\pi R}{\lambda} \Delta \theta \right) \right| \approx 0$, we choose the sampling points based on the first zero of $J_0(\cdot)$, denoted by $j_{0,1}$. Let $\alpha = j_{0,1}$. The elevation angle sampling grid is then:
\begin{equation}  \label{eq10} % Equation number kept same as original
    \theta _ { t } = \sin ^ { - 1 } \left( t \cdot \frac { \lambda \alpha } { 2 \pi R } \right) , \quad t = 0 , 1,\cdots , T,
    \end{equation}
where $T=\left\lfloor \frac { 2 \pi R } { \lambda \alpha } \right\rfloor$, and $\lfloor \cdot \rfloor$ is the floor function. 

\subsubsection{Sampling Method of Azimuth Angle}
Similarly, for azimuth angle sampling, we fix the distance and elevation angle ($r_p = r_q = r$, $\theta_p = \theta_q = \theta$) and consider different azimuth angles $\phi_p \neq \phi_q$. The correlation can be approximated as:
\begin{equation}   \label{eq11} % Equation number kept same as original
  \begin{split}
  &f(r, \theta, \phi_p; r, \theta, \phi_q) \approx |\mathbf{a}_\mathrm{S}(\theta_p, \phi)^H \mathbf{a}_\mathrm{S}(\theta_q, \phi)| \\
  &\overset{(d)}{=} \frac{1}{N} \left| \sum_{n=1}^{N} e^{j\frac{4\pi}{\lambda} R \sin \theta \sin \left( \frac{\phi_p - \phi_q}{2} \right) \cos \left( \frac{\pi}{2} - \frac{\phi_p + \phi_q - 2\varphi_n}{2} \right)} \right| \\
  &\approx \left| J_0 \left( \frac{4\pi R \sin \theta}{\lambda} \Delta \phi \right) \right|,
  \end{split}
  \end{equation}
 where step $(d)$ is obtained using the sum-to-product formula $\cos \alpha - \cos \beta=-2 \sin \frac{\alpha + \beta}{2} \sin \frac{\alpha - \beta}{2}$ and the trigonometric transformation formula $\sin \alpha=\cos \left( \frac{\pi}{2} - \alpha \right)$. The subsequent approximation is similar to that of \eqref{eq9}. Setting $\left| J_0 \left( \frac{4 \pi R \sin \theta}{\lambda} \Delta \phi \right) \right| = 0$, the azimuth sampling grid is 
  \begin{equation}  \label{eq12} % Equation number kept same as original
    \phi_s = s \cdot 2\sin^{-1} \left( \frac{\lambda \alpha}{4\pi R \sin\theta} \right),\quad s=0,1,\cdots, S,
    \end{equation}
    where $S=\left\lfloor  \pi / \sin^{-1} \left( \frac{\lambda \alpha}{4 \pi R \sin \theta} \right)\right\rfloor$.

\subsubsection{Sampling Method of Distance}
Similar to the angle sampling methods, we consider sampling two near-field steering vectors associated with the same angles $(\theta_p=\theta_q = \theta, \phi_p=\phi_q = \phi)$ but different distances $(r_p \neq r_q)$. The corresponding codebook column correlation $f ( r _ { p } , \theta , \phi , r _ { q } , \theta , \phi )$ can be approximated as \cite{re4}
\begin{equation}
  f ( r _ { p } , \theta , \phi ; r _ { q } , \theta , \phi ) \approx | J _ { 0 } ( \beta ) |,
  \end{equation}
  where $\beta = \frac{\pi R^2 \sin^2 \theta}{2\lambda} \left| \frac{1}{r_p} - \frac{1}{r_q} \right|$.
  Since the function $|J_0(x)|$ is monotonically decreasing for $x$ within the interval $(0,\alpha)$, the condition $|J_0(\beta)| \leq \Delta$ can be approximately satisfied by ensuring that $\beta \geq \beta_\Delta$, where $\beta_\Delta$ is chosen such that $|J_0(\beta_\Delta)| = \Delta$.
  Therefore, by controlling the sampling precision according to this principle, the correlation between steering vectors focused at the same angle but different distances can be maintained below the desired threshold $\Delta$.
  Finally, an effective distance sampling method obtained from this consideration is given as
  \begin{equation}
    \label{eq14} % Ensure this label matches the original if needed
    r_z = \frac{1}{z} Z_\Delta\sin^2 \theta, \quad z = 0, 1, 2, \cdots ,
    \end{equation}
    where $Z_\Delta = \frac{\pi R^2}{2\lambda\beta_\Delta}$.
    This approach leads to non-uniform sampling of the distance $r$ over the range $(r_{\min}, + \infty)$, with the sampling density depending on the elevation angle $\theta$. Here, $r_{\min}$ denotes the minimum communication distance of interest, where $r_{\rm{min}}  > 0.5 \sqrt{{D^3}/{\lambda}} $, with $D$ representing the array aperture. The overall design process of the spherical domain transformation matrix codebook incorporating these sampling methods is summarized in \textbf{Algorithm 1}. 
\begin{algorithm}
  \caption{Proposed Spherical-Domain Transform Matrix Codebook $\mathbf{W}$ Design Algorithm}
  \label{alg:codebook_design}
  \textbf{Input:} $r_{\text{min}}$, $N$, $R$, $\lambda$, $\Delta$. \\
   \textbf{Output:} $\mathbf{W}$.
  \begin{algorithmic}[1]
  \STATE Initialize $\mathbf{W} = []$. Let $\alpha = j_{0,1}$. Determine $\beta_\Delta$ from $|J_0(\beta_\Delta)| = \Delta$.
  \STATE Determine $\theta_t$ via \eqref{eq10}.
  \FOR{each valid elevation angle $\theta_t$}
      \STATE Determine $\phi_s$ via \eqref{eq12} for current $\theta_t$.
      \STATE Determine $r_z$ via \eqref{eq14} for current $\theta_t$, ensuring $r_z \ge r_{\min}$.
      \FOR{each valid azimuth angle $\phi_s$}
          \FOR{each valid distance $r_z$}
              %\STATE Construct steering vector $\mathbf{b}_\mathrm{S}(r_z, \theta_t, \phi_s)$ using \eqref{eq4} and \eqref{eq5}.
              \STATE Append $\mathbf{b}_\mathrm{S}(r_z, \theta_t, \phi_s)$ as a column to $\mathbf{W}$.
          \ENDFOR
      \ENDFOR
  \ENDFOR
  %\STATE Let $G$ be the total number of columns in $\mathbf{W}$.
  \RETURN $\mathbf{W}$
  \end{algorithmic}
\end{algorithm}

\begin{figure}[t]
  \centering
  
  \subfloat[sampling points for $z=1$]{
      \includegraphics[width=0.3\textwidth]{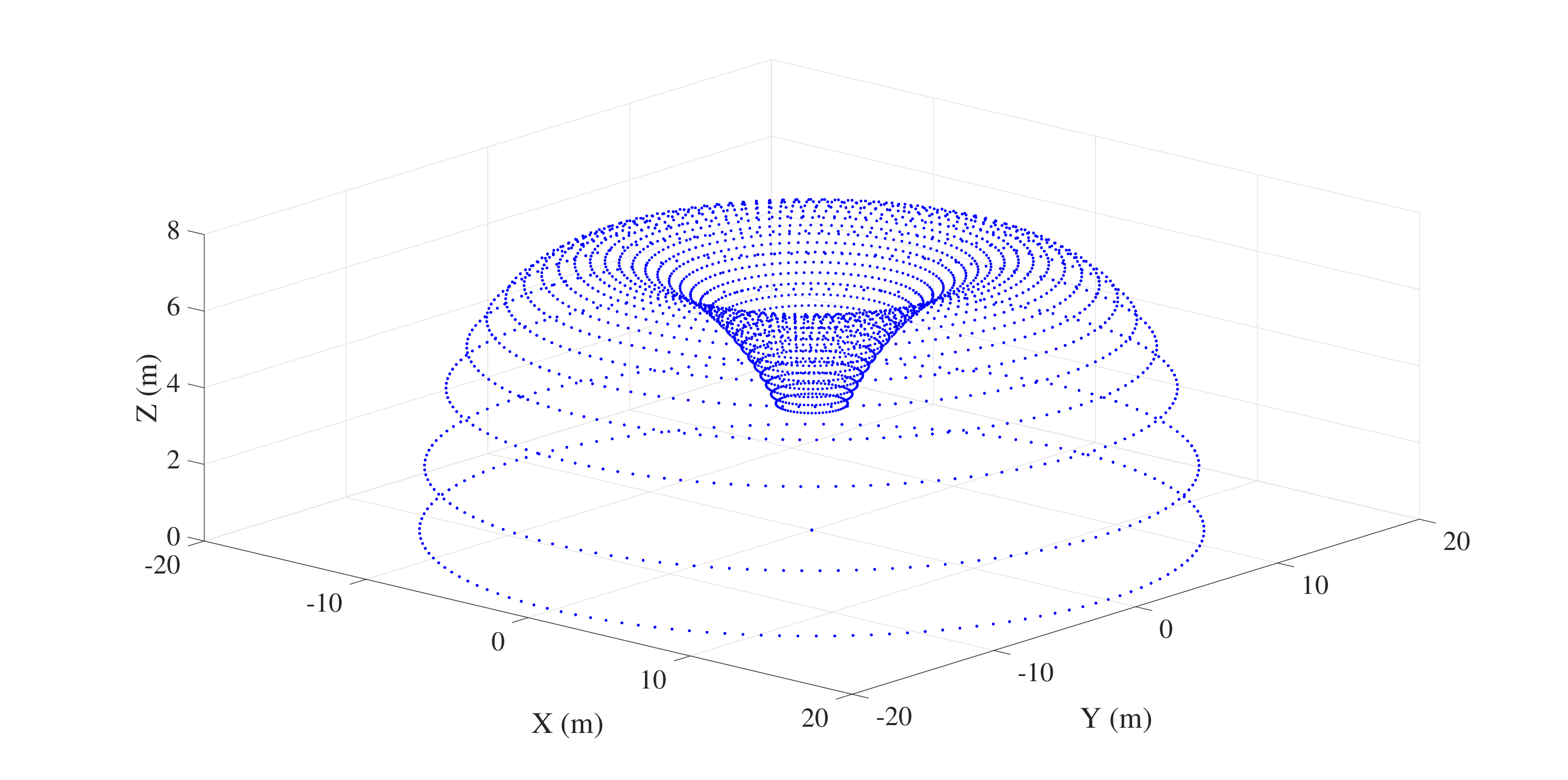} 
      \label{fig:subfig1}
  }

  \subfloat[sampling points for $\theta=\pi/2$]{
      \includegraphics[width=0.2\textwidth]{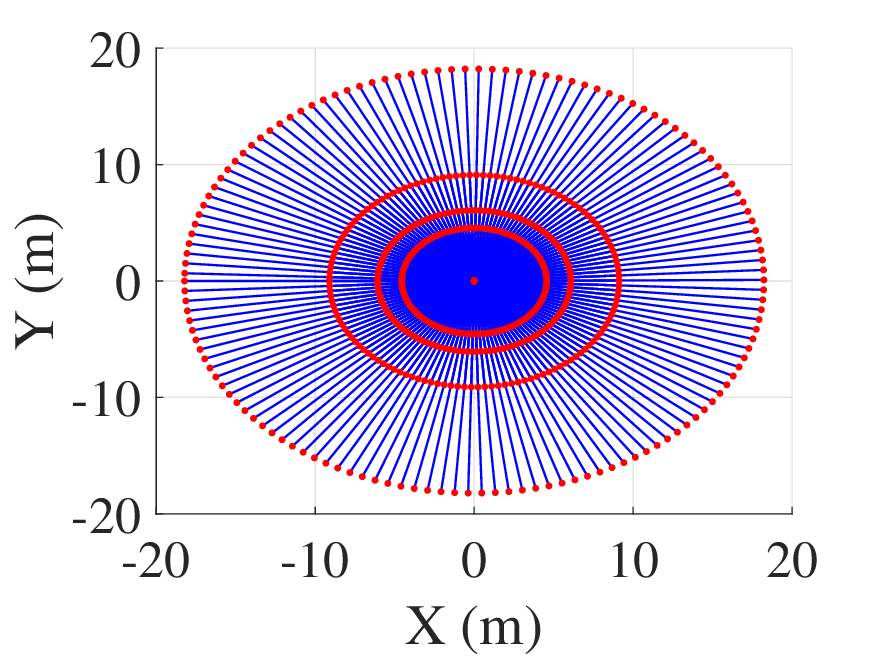} 
      \label{fig:subfig2}
  } \subfloat[sampling points for $\phi=0$]{
      \includegraphics[width=0.2\textwidth]{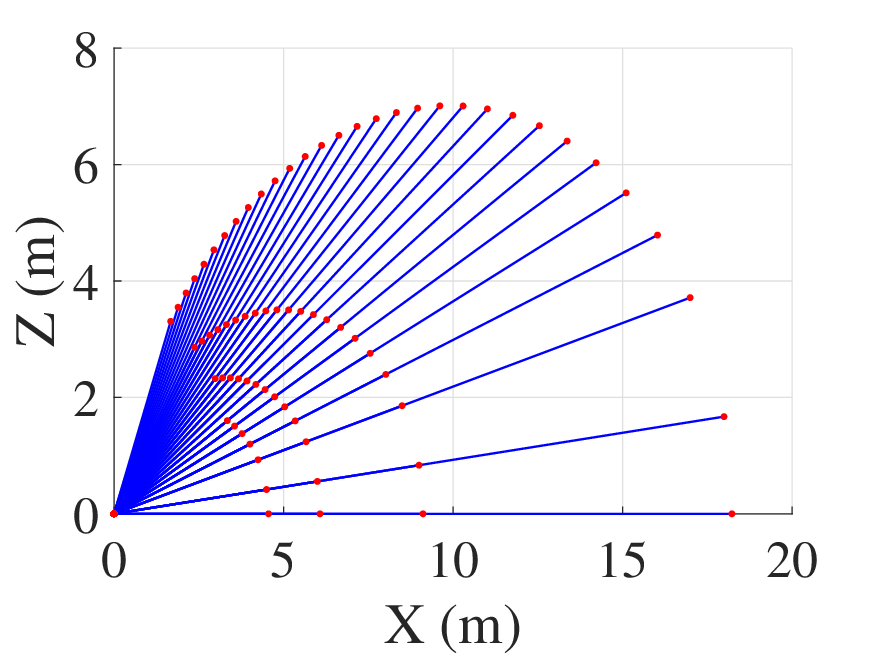} 
      \label{fig:subfig3}
  }\caption{Example spherical-domain codebook sampling points. (a) Distance domain samples (fixed angle, $z$). (b) Azimuth angles/points (fixed $r, \theta$). (c) Elevation angles/points (fixed $r, \phi$).  }
  \vspace{-0.25cm}
  \label{figure2}
\end{figure}

Fig. \ref{figure2} visually illustrates examples of the sampling points generated by this procedure for specific parameter choices. Since the near-field region becomes negligible at small elevation angles\cite{re4}, we adopt far-field communication as the default assumption. Compared to conventional 2D polar-domain codebooks (distance/azimuth only), our design incorporates elevation sampling. Critically, by exploiting UCA rotational symmetry and deriving non-uniform azimuth sampling intervals from Bessel function zeros, our approach overcomes the inherent azimuth blind zones associated with ULAs. Furthermore, the analytical derivation of sampling intervals systematically minimizes the cross-correlation between steering vectors at different angles, directly enhancing sparse recovery performance.

\subsection{S-SOMP-Based Channel Estimation Scheme}
Leveraging the codebook $\mathbf{W}$, we propose a grid-based S-SOMP channel estimation scheme.
Specifically, the spherical-domain channel representation $\mathbf{h}^\mathrm{S}_m$ in \eqref{eq6} exhibits sparsity. Since the physical paths are common across subcarriers, the sparsity support sets of the spherical-domain channel vectors are identical across different subcarrier frequencies $f_m$. This joint sparsity structure enables the use of the SOMP algorithm.
Therefore, the spherical-domain channel estimation can be formulated as a joint sparse signal recovery problem.

Exploiting this common support allows for simultaneous estimation across subcarriers, thereby enhancing accuracy. We can rewrite \eqref{eq2} in a matrix form encompassing all $M$ subcarriers as
\begin{equation}
  \mathbf{Y}=\mathbf{AW} \mathbf{H}^\mathrm{S} + \mathbf{N},
\end{equation}
where $\mathbf{Y} = [\mathbf{y}_1, \mathbf{y}_2, \dots, \mathbf{y}_M]$ and $\mathbf{N} = [\mathbf{n}_1, \mathbf{n}_2, \dots, \mathbf{n}_M]$.

The proposed S-SOMP algorithm is detailed as follows and summarized in \textbf{Algorithm 2}.
Step 1 initializes the residual matrix $\mathbf{R = Y}$ and the support set $\mathbf{\Omega} = \emptyset$.
Step 3 calculates the correlation matrix $\mathbf{\Gamma} = (\mathbf{AW})^{H}\mathbf{R}$, capturing the correlation between each column of the dictionary matrix $\mathbf{AW}$ and the current residual matrix $\mathbf{R}$.
Step 4 identifies the index $p^*$ corresponding to the spherical-domain sampling point with the highest correlation energy summed across all subcarriers: $p^{*} = \arg\max_{p}\sum_{m=1}^{M}|\mathbf{\Gamma}(p,m)|^{2}$. This greedily selects the most significant path component in the current iteration.
Step 5 adds the identified index $p^*$ to the support set: $\mathbf{\Omega} = \mathbf{\Omega} \cup \{p^{*}\}$.
Step 6 computes the least squares estimate of the spherical-domain channel coefficients $\widehat{\mathbf{H}}_{\mathbf{\Omega}, :}^\mathrm{S}$ restricted to the support set $\mathbf{\Omega}$: $\widehat{\mathbf{H}}_{\mathbf{\Omega}, :}^\mathrm{S} = ((\mathbf{AW})_{:, \mathbf{\Omega}})^{\dagger}\mathbf{Y}$.
Step 7 updates the residual matrix by removing the contribution of the estimated components: $\mathbf{R} = \mathbf{R} - (\mathbf{AW})_{:, \mathbf{\Omega}}\widehat{\mathbf{H}}_{\mathbf{\Omega}, :}^\mathrm{S}$.
These steps (3-7) are iterated $\widehat{L}$ times to detect all path components.
Finally, the estimated near-field channel $\widehat{\mathbf{H}}$ is reconstructed using the estimated sparse coefficients $\widehat{\mathbf{H}}_{\mathbf{\Omega},:}^\mathrm{S}$ and the corresponding columns $\mathbf{W}_{:, \mathbf{\Omega}}$ from the codebook: $\widehat{\mathbf{H}} = \mathbf{W}_{:, \mathbf{\Omega}}\widehat{\mathbf{H}}_{\mathbf{\Omega}, :}^\mathrm{S}$.

\begin{algorithm}
\caption{Proposed S-SOMP Algorithm}
\label{alg:spherical_somp}
\textbf{Input:} $\mathbf{Y}$, $\mathbf{A}$, $\mathbf{W}$, $\widehat{L}$. \\
\textbf{Output:} $\widehat{\mathbf{H}}$.
 \begin{algorithmic}[1]
\STATE Initialization: $\mathbf{R = Y}$, $\mathbf{\Omega} = \emptyset$
\FOR{$l \in \{1, 2, \dots, \widehat{L}\}$}
    \STATE Calculate the correlation matrix: $\mathbf{\Gamma} = (\mathbf{AW})^{H}\mathbf{R}$.
    \STATE Detect new support: $p^{*}=\arg\max_{p}\sum_{m=1}^{M}|\mathbf{\Gamma}(p,m)|^{2}$.
    \STATE Update support set: $\mathbf{\Omega} = \mathbf{\Omega} \cup \{p^{*}\}$.
    \STATE Orthogonal projection: $\widehat{\mathbf{H}}_{\mathbf{\Omega}, :}^\mathrm{S} = ((\mathbf{AW})_{:, \mathbf{\Omega}})^{\dagger}\mathbf{Y}$.
    \STATE Update residual matrix: $\mathbf{R} = \mathbf{R} - (\mathbf{AW})_{:, \mathbf{\Omega}}\widehat{\mathbf{H}}_{\mathbf{\Omega}, :}^\mathrm{S}$.
\ENDFOR
\STATE Reconstruct channel: $\widehat{\mathbf{H}} = \mathbf{W}_{:, \mathbf{\Omega}}\widehat{\mathbf{H}}_{\mathbf{\Omega}, :}^\mathrm{S}$
\RETURN $\widehat{\mathbf{H}}$
\end{algorithmic}
\end{algorithm}

Next, we analyze the computational complexity of the proposed algorithm.
The overall complexity is primarily determined by the iterative process (Steps 2-8 in \textbf{Algorithm 2}).
Given the matrix dimensions $\mathbf{A} \in \mathbb{C}^{PN_{\mathrm{RF}} \times N}$, $\mathbf{W} \in \mathbb{C}^{N \times G}$, $\mathbf{R} \in \mathbb{C}^{PN_{\mathrm{RF}} \times M}$, and $\mathbf{Y} \in \mathbb{C}^{PN_{\mathrm{RF}} \times M}$,
the computational complexities for steps 3-7 within the $l$-th iteration ($l=1, \dots, \widehat{L}$) are $\mathcal{O}(PN_{\mathrm{RF}}NGM)$, $\mathcal{O}(GM)$, $\mathcal{O}(1)$, $\mathcal{O}(l^2PN_{\mathrm{RF}} + lPN_{\mathrm{RF}}M)$, and $\mathcal{O}(l PN_{\mathrm{RF}}M)$, respectively.
Accumulating the complexity over $\widehat{L}$ iterations yields an overall complexity of approximately $\mathcal{O}(\widehat{L}PN_{\mathrm{RF}}NGM + \widehat{L}GM + \widehat{L}^3PN_{\mathrm{RF}} + \widehat{L}^2PN_{\mathrm{RF}}M)$. The final channel reconstruction (Step 9) adds a complexity of $\mathcal{O}(N\widehat{L}M)$.
Since the number of paths $\widehat{L}$ is typically small in practical scenarios, the computational complexity of the proposed method is manageable.

\section{Simulation Results}
In this section, the performance of the proposed channel estimation scheme is evaluated through numerical simulations. Unless otherwise specified, the simulation parameters are set as follows: $f_c = 30$ GHz, $N = 512$, $K = 4$, $d = 0.005$ m, and $\Delta = 0.55$. In this setup, the Rayleigh distance is 132.9 m. Users are randomly located within the range of elevation angle $\theta \in \left( 0, \frac{\pi}{2} \right)$, azimuth angle $\boldsymbol{\phi} \in \left(-\frac{\pi}{2}, \frac{\pi}{2}\right)$, and distance $r \in (4\,\text{m}, 25\,\text{m})$. The normalized mean square error (NMSE), defined as $\mathrm{NMSE} = \mathbb{E} \left( \frac{\|\mathbf{H} - \hat{\mathbf{H}}\|_F^2}{\|\mathbf{H}\|_F^2} \right)$, is used as the performance metric. The benchmark schemes for comparison are:
\begin{itemize}
    \vspace{-0.1cm}
    \item SOMP \cite{re6}: It transforms the channel $\mathbf{h}_m$ into the angular-domain (azimuth angle) using a spatial Fourier transform matrix codebook and employs the OMP algorithm to jointly estimate channels across multiple subcarriers.
    \item P-SOMP \cite{re7}: It constructs a polar-domain (distance and azimuth angle) sparse representation of $\mathbf{h}_m$ via a polar transform codebook and subsequently performs channel estimation using the SOMP algorithm.
    \item LS: The least squares (LS) estimator minimizes the mean square error (MSE) between the received signal and the model prediction, given by $\mathbf{\hat{H}} = \underset{\mathbf{H}}{\mathrm{argmin}} \|\mathbf{Y} - \mathbf{A}\mathbf{H}\|_F^2$. 
    \item CRLB: The Cramér-Rao lower bound (CRLB) provides the theoretical minimum NMSE, Assuming complete knowledge of all users' and scatterers' positions is available at step 5 of \textbf{Algorithm 2}, channel estimation is then performed in steps 6 and 8.
  \vspace{-0.1cm}
  \end{itemize}

\begin{figure}[t]
  \centering
  \includegraphics[scale=0.4]{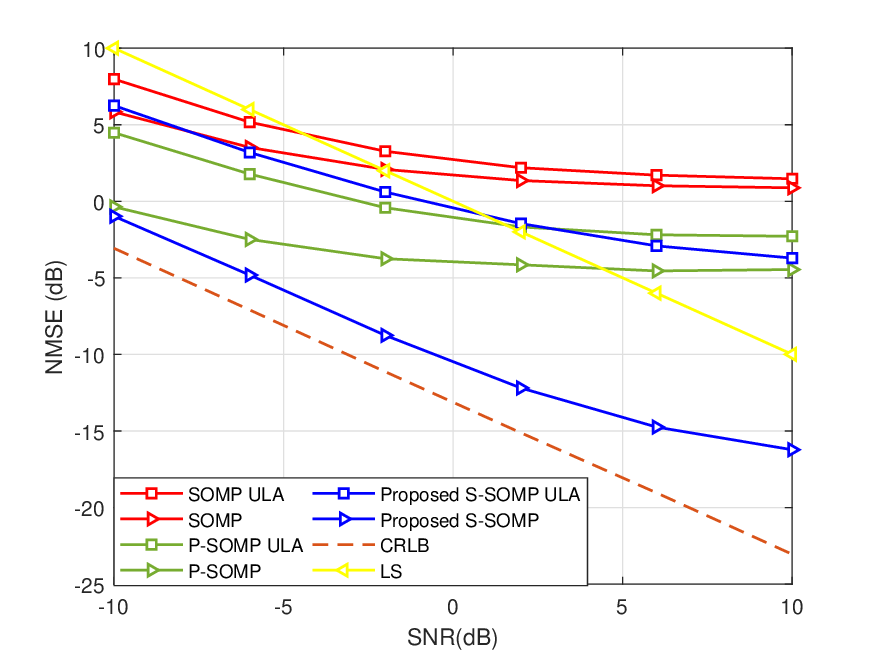}
  \caption{NMSE of different schemes versus SNR.}
  \vspace{-0.25cm}
  \label{figure3}
\end{figure}

Fig. \ref{figure3} illustrates the NMSE performance of different schemes versus the signal-to-noise ratio (SNR), where $\mathrm{SNR} = \mathbb{E} \left( \frac{\|\mathbf{H} \|_F^2}{\|\mathbf{N}\|_F^2} \right)$ \cite{re7} and the pilot length is $P = 32$.
% Removed the sentence about Genie-aided LS, as CRLB is defined above as the theoretical bound. If you plot a Genie-aided result *as* the CRLB, you might add "(calculated assuming perfect path knowledge)" after CRLB in the legend/text.
For comparison, we also evaluate the performance of ULA configurations with the same number of antennas using corresponding algorithms (SOMP ULA, P-SOMP ULA, and proposed S-SOMP ULA).
It is observed that the NMSE decreases for all schemes as SNR increases.
Notably, the proposed S-SOMP method significantly outperforms all benchmark algorithms, especially in the high SNR regime.
Furthermore, comparing array geometries, the NMSE performance with UCA consistently surpasses that with ULA for all applicable algorithms.
This advantage stems from the proposed spherical-domain transformation codebook $\mathbf{W}$ used with S-SOMP, which enables a comprehensive channel characterization by explicitly sampling the three-dimensional spatial parameters: distance, azimuth, and elevation.
The UCA geometry supports this 3D characterization, allowing it to capture spatial channel features more accurately.
In contrast, the ULA model inherently struggles in 3D scenarios; its linear structure typically resolves only distance and azimuth, effectively simplifying elevation information (e.g., assuming $\theta = \pi/2$). % Adjusted explanation for clarity and accuracy.
The proposed S-SOMP algorithm's performance closely approaches the CRLB, demonstrating its near-optimal effectiveness.

\begin{figure}[t]
  \centering
  \includegraphics[scale=0.4]{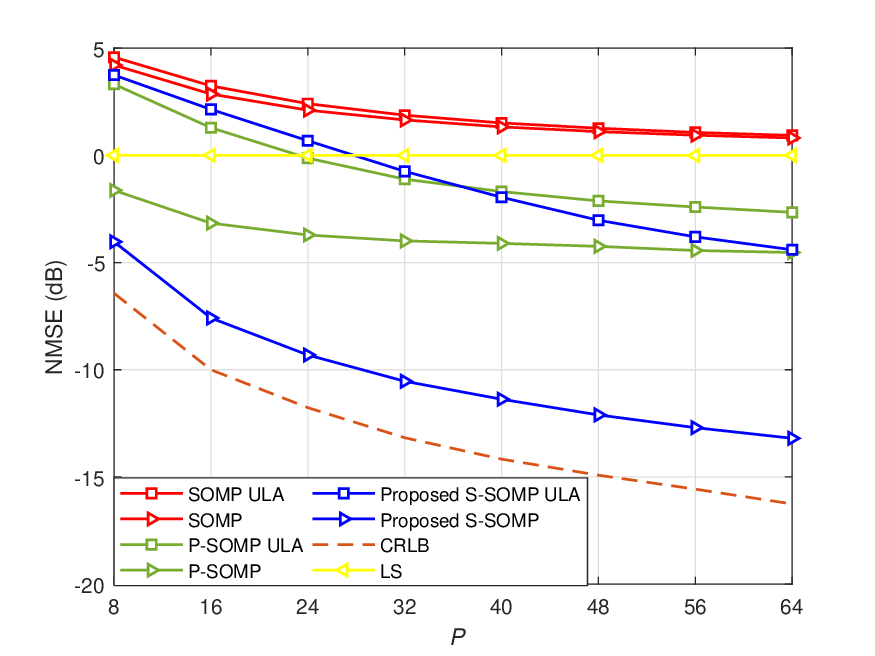}
  \caption{NMSE of different schemes versus pilot length $P$.} % Corrected typo "poilt"
  \vspace{-0.25cm}
  \label{figure4}
\end{figure}

Fig. \ref{figure4} depicts the NMSE versus the pilot length $P$, varying from 8 to 64, at a fixed SNR of 5 dB.
The results show that increasing the pilot length improves the NMSE for all schemes, as expected.
However, the proposed S-SOMP method exhibits a steeper improvement in channel estimation accuracy and consistently maintains superior performance across different pilot lengths.
The reason is that the low cross correlation codebook design controls the sampling interval by using the zeros of the Bessel function, thereby significantly improving the stability of sparse recovery.
This indicates that the proposed method can potentially reduce the pilot overhead required for accurate near-field channel estimation in XL-MIMO systems.

\section{Conclusion}
In this letter, we proposed an S-SOMP scheme to address the 3D near-field channel estimation challenge in XL-MIMO systems with UCAs. Its core is a spherical-domain transform codebook jointly modeling distance, azimuth, and elevation angles. Orthogonality constraints incorporated into the codebook design ensure low steering vector cross-correlation, enhancing sparse signal recovery performance. Simulations showed that the proposed S-SOMP scheme, particularly with a UCA, achieves substantially lower estimation errors compared to conventional angular/polar-domain algorithms and ULA configurations. Furthermore, using the UCA geometry with our proposed estimator facilitates accurate channel estimation with potentially reduced pilot overhead.

\end{document}